# Physics Archives

August 2010

## Towards Solving the Inverse Protein Folding Problem

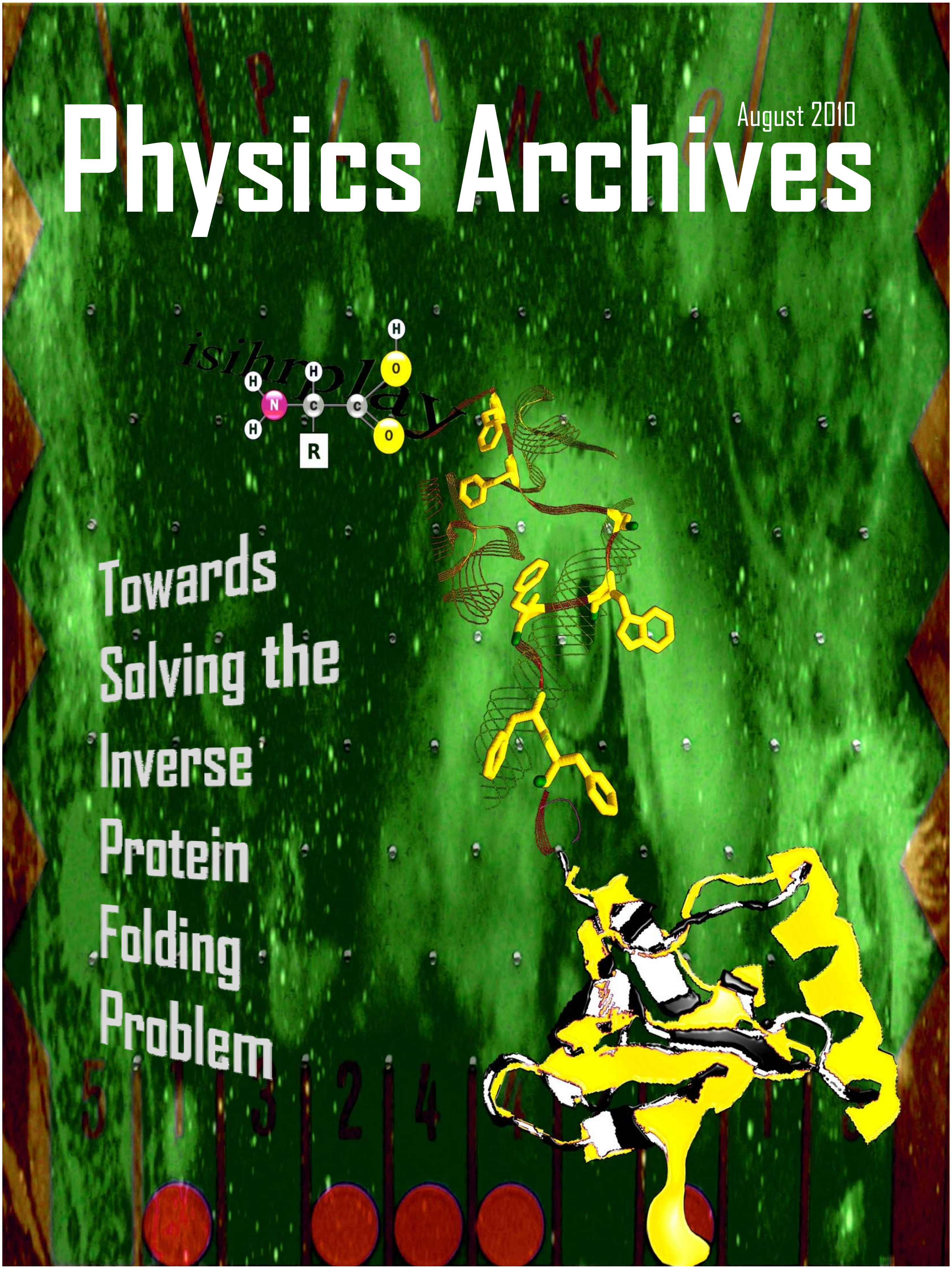

**Towards Solving the Inverse Protein Folding Problem**

**Yoojin Hong[1,2], Kyung Dae Ko[2,3], Gaurav Bhardwaj[2,3], Zhenhai Zhang[2,4], Damian B. van Rossum[2,3]*, and Randen L. Patterson[2,3]***

[1] Department of Computer Science and Engineering, The Pennsylvania State University, USA

[2] Center for Computational Proteomics, The Pennsylvania State University, USA

[3] Department of Biology, The Pennsylvania State University, USA

[4] Department of Biochemistry and Molecular Biology, The Pennsylvania State University, USA

[*] To whom correspondence should be addressed: rlp25@psu.edu; dbv10@psu.edu

Running title "Fold-recognition in the twilight zone"

## ABSTRACT

**Accurately assigning folds for divergent protein sequences is a major obstacle to structural studies and underlies the inverse protein folding problem. Herein, we outline our theories for fold-recognition in the "twilight-zone" of sequence similarity (<25% identity). Our analyses demonstrate that *structural sequence profiles* built using Position-Specific Scoring Matrices (PSSMs) significantly outperform multiple popular homology-modeling algorithms for relating and predicting structures given only their amino acid sequences. Importantly, *structural sequence profiles* reconstitute SCOP fold classifications in control and test datasets. Results from our experiments suggest that *structural sequence profiles* can be used to rapidly annotate protein folds at proteomic scales. We propose that encoding the entire Protein DataBank (~1070 folds) into *structural sequence profiles* would extract interoperable information capable of improving most if not all methods of structural modeling.**

## INTRODUCTION

It has been proposed that the number of distinct native state protein folds is extremely limited(1). In addition, structure is more conserved than sequence similarity(1-3). Taken together, these attributes underscore the inverse protein folding problem; whereby the vast and varied numbers of primary amino acid sequences that exist in biology occupy a relatively limited number of structural folds. Due to the extreme divergence (≤25% pairwise identity) that can exist between structurally resolved (template) sequences and structurally unknown (target) sequences, fold-classification is often compromised. Thus, the crucial information specifying protein structure must be contained in a very small fraction of the amino acid sequence, making the informative points hard to measure. Therefore, a solution to the inverse protein folding problem must be able to identify these information points and use them to relate targets to appropriate template sequences.

From a practical standpoint, the inverse protein folding problem manifests at various stages of structural modeling, depending on the method employed. For example, in homology-based threading techniques (e.g. Swiss-Model, Modeller, etc(4, 5)), unless the target is aligned to a structure of correct fold initially, the model will be inaccurate. For approaches that use physical modeling techniques (and hybrids thereof), as the fold is constructed from smaller structural units, they can often be assembled in multiple low-energy conformations. Mistakes can occur during the assembly of the fold using these approaches; further, they are often computationally expensive. The inverse protein folding problem has also been attacked using profiling methods such as FFAS03, SAM-T2K, and prof_sim(6-8). However, all of these algorithms experience a "glass-ceiling" whereby at relevant statistical limits, only <30% of benchmark structural datasets can be properly classified between sequences of ≤25% identity. Thus, if the correct fold for a given target sequence could be rapidly and reliably defined, most if not all structural modeling approaches could be improved.

We recently theorized that a computational platform could be developed for sensitive homology detection and secondary structure annotation using rps-BLAST compatible PSSM libraries (9-11). In the present manuscript, we report that *structural sequence profiles* (i.e. fold-specific PSSM libraries) are a robust method of fold-classification which works in the "twilight-zone" of sequence similarity using simple algebra. Our findings demonstrate that *structural sequence profiles* are a new performance benchmark for the detection of distant structural homology. These results also provide support for our theories that sufficiently large PSSM libraries provide a solution to the inverse protein folding problem.

## RESULTS

### *Fold-based Structural Sequence Profiles*

The power behind our *structural sequence profiles* is derived from libraries of Position-

Specific Scoring Matrices (PSSMs, i.e. profile) of functionally or structurally similar proteins, which contain a frequency table for substitutions that occur in related sequences; PSSMs are a powerful measure of homology. Indeed, it is well-established that PSSMs contain more information than individual sequences(12-14). We take advantage of the increased information content of PSSMs and quantify their alignments within a *structural sequence profile*. There are three features which make our method distinct from traditional sequence analysis methods. First, we measure targets with multiple structure-specific PSSM libraries. Second, we quantify low identity alignments, which are traditionally considered statistically insignificant. Third, we consider all relationships (to the same fold and different folds) to extract meaningful signals, which appear to be important for measurements in the "twilight zone"(9-11, 15).

### *Alignment Comparisons and Information Content*

To test our method, we used the TZ-SABmark which is a carefully curated set of fold-specific sequences of remote homology(16). Each fold-specific sequence group represents a SCOP(17) fold classification of related sequences with ≤25% sequence identity. From the original TZ-SABmark, 534 sequences from 61 fold groups (avg. length of $135.27 \pm 89.39$ s.d.) selected at random were used as a test set.

Out method involves three steps to infer remote structural homology between proteins (Fig. 1). First, we collect the sequences for each of the 61 test fold groups from the Protein Data Bank (PDB(18)). PDB sequences in the TZ-SABmark were excluded to avoid debate. Except for one fold group (SCOP fold b.1; *Immunoglobulin-like beta-sandwich* fold) which already has >1000 PDB sequences, the PDB sequences of all 60 fold groups were expanded by PSI-BLAST(13) search against NCBI NR database using themselves as targets. The sequences similar to the PDB sequences (≥90% identity) were removed. For each fold group, redundant or highly similar sequences (≥40% identity) were also eliminated. Fold-specific libraries for 61 fold groups were then built by generating PSSMs from the sequences obtained from PSI-BLAST. Following, fold-specific PSSMs were compiled as an rps-BLAST(19) compatible database (Fig. 1b).

Second, each query sequence is then searched against the 61 fold-specific PSSM libraries using rps-BLAST. The alignments returned from the search are filtered out if they do not satisfy our e-value and coverage thresholds (i.e., alignment length as a function of library PSSM length). Third, given the alignments to a fold-specific library, a fold-specific score is calculated and encoded in *structural sequence profile* where each query is a vector of fold-specific scores (Fig. 1c, see Methods for more details). As a quantitative measure of how two targets are similar (i.e. the *structural similarity score*), we calculated Pearson's correlation coefficient between their vectors. In the studies, alignments encoded in *structural sequence profiles* were collected using either of e-value 0.01, no coverage or e-value $10^{10}$, 80% coverage thresholds.

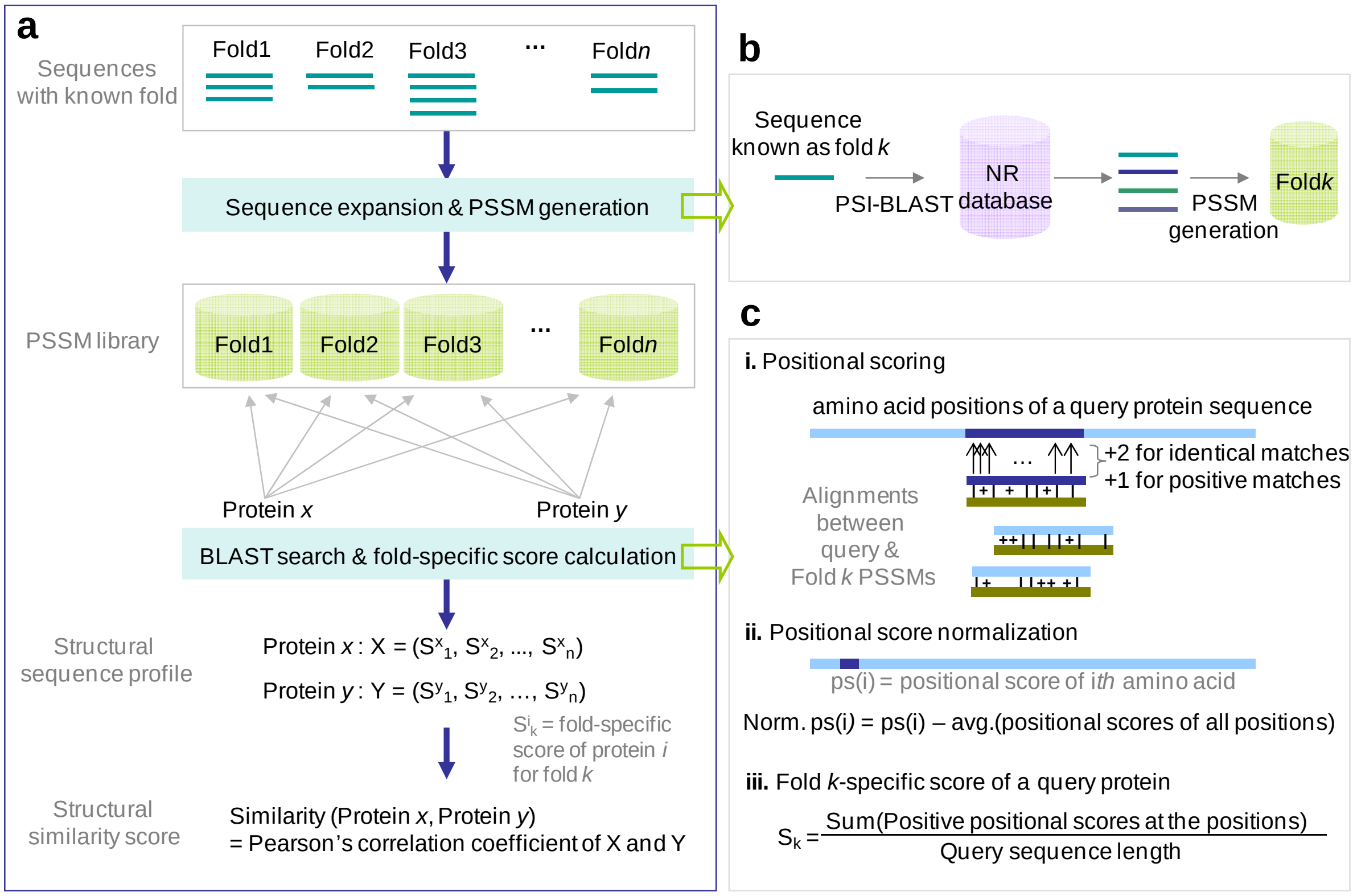

**Figure 1. Computational Pipeline. (a)** Basic pipeline for generating *structural sequence profiles*. For each fold, structurally resolved sequences are collected, expanded by PSI-BLAST, and used to generate PSSMs to create a fold-specific library. Each fold-specific library, compiled as rps-BLAST database, can be searched at varying e-value thresholds. Given the alignments returned after filtering by % coverage, a fold-specific score for the query is calculated. By repeating this process using different fold-specific libraries, the query protein can be represented as a *structural sequence profile*, which is a vector of fold-specific scores. To calculate *structural similarity score* of two proteins, Pearson's correlation coefficient of their *structural sequence profiles* is calculated. **(b)** PDB sequences with known fold are collected, and expanded by PSI-BLAST search against NCBI NR database with each of the PDB sequences as a query. After removing redundant or highly similar sequences, PSSMs are generated from the collected sequences by PSI-BLAST for a fold-specific library. **(c)** Given an alignment between a query and a fold-specific PSSM, each query amino acid is scored +2 for identical matches and +1 for conserved matches. After scoring with all alignments against PSSMs from the fold-specific library, positional scores of query amino acids were normalized by subtracting the average positional scores. The fold-specific score of a protein is calculated by dividing the sum of all positive positional scores by a query sequence length.

We first evaluated sequence similarity between TZ-SABmark test sequences and the sequences used for building fold-specific libraries. Figure 2a plots cumulative frequency distributions of pairwise %identity between pairs of TZ-SABmark test sequences and PSSMs from their true- and false-fold groups. These statistics demonstrate that ~95% of all same-fold pairs have <20% pairwise identity. Indeed, this distribution is negligibly distinct from comparisons of different-fold pairs. Additionally, we compared the sequence similarity between the PDB reference sequences and those PSSMs which were obtained through their PSI-BLAST expansion. PSSMs used to define fold-specific libraries are also in the "twilight zone". Taken together, this indicates that our information source is derived from low-identity alignments and not merely from redundancy.

It is reasonable to consider that a protein would have a larger score for its true-fold than for

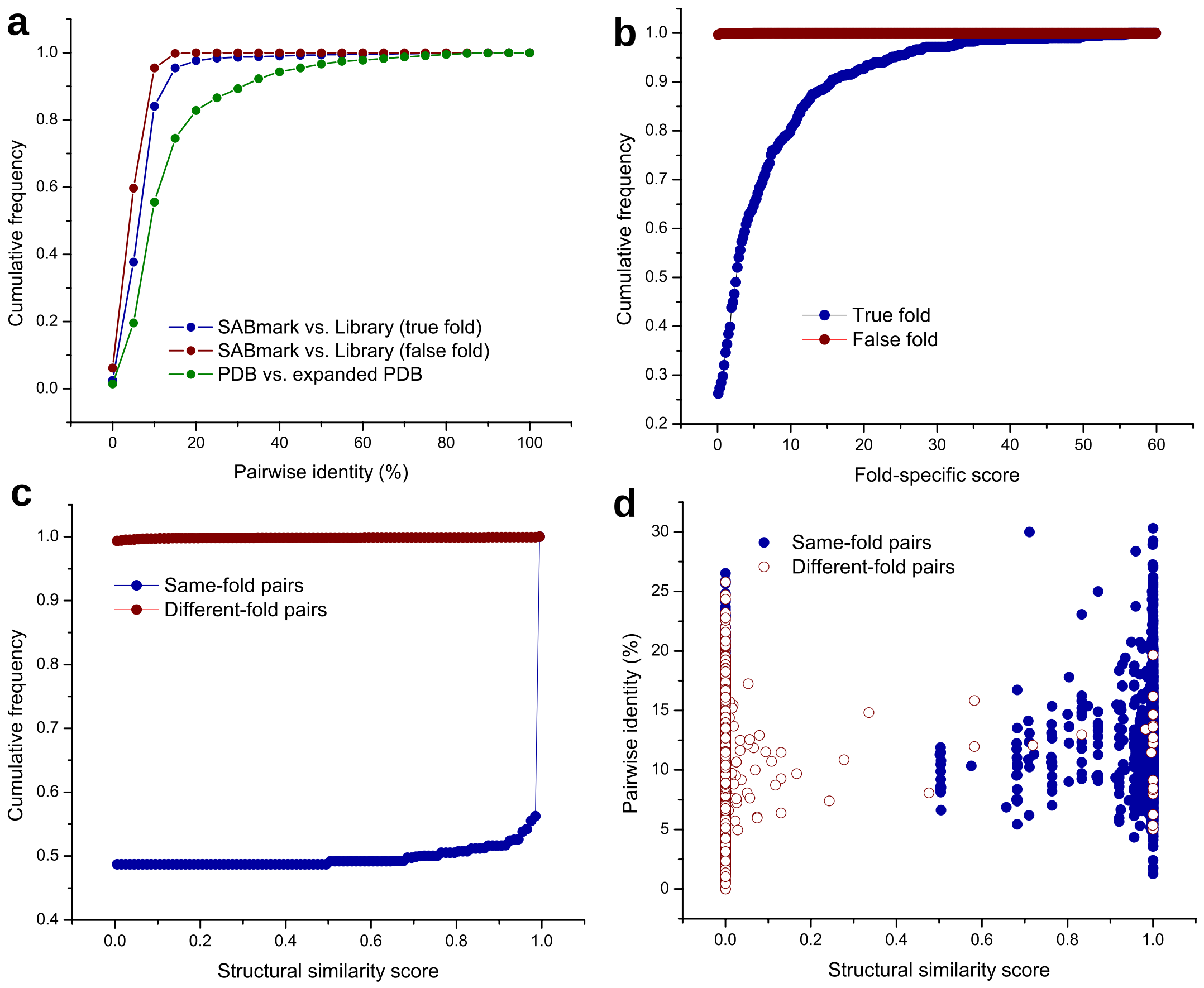

**Figure 2. Characterization of Structural Similarity Scores. (a)** The distribution of % pairwise identity between pairs of TZ-SABmark sequences and the library sequences of the same-fold (blue) and different-folds (red) and % pairwise identity between the original PDB sequences and PSI-BLAST expanded sequences (green). All three comparisons demonstrate that nearly all of the sequence alignments reside in the "twilight-zone" of sequence similarity. The pairwise identity was calculated from Needleman-Wunsch global alignments(22) with BLOSUM62(23), Gap opening penalty 10, and Gap extension penalty 0.5. **(b)** The distributions of query sequence scores for each fold-specific library. **(c)** Cumulative frequencies of the *structural similarity scores* between pairs of same-fold (blue) and different-fold (red) query sequences. For this measurement, 3,428 same-fold pairs and 65,536 different-fold pairs were measured from 534 sequences. **(d)** *Structural similarity scores* between pairs of same-fold and different-fold query sequences were plotted versus their % pairwise sequence identity. This data shows an independent trend between the *structural similarity score* and pairwise identity in the "twilight-zone" of sequence similarity. For different-fold pairs, randomly selected 10,000 data points were plotted. The statistics in the Fig. b,c,d were obtained given the setting of e-value 0.01, no coverage threshold.

its false-folds; this is confirmed in Figure 2b and demonstrates that our fold libraries are specific. We observe that 99.6% of the query sequences have fold-specific scores $\leq 0.1$ for different-folds, while only 25.7% of them have scores $\leq 0.1$ for same-folds. Given these data, if we annotate each protein by the highest fold-specific score, the folds of 74.2% of TZ-SABmark test sequences can be predicted correctly (e-value 0.01, no coverage). Figure 2c shows cumulative frequencies of structural similarity score between pairs of same-fold (blue, 3,428 pairs) and different-fold (red, 65,536 pairs) query sequences. ~58.2% of same-fold pairs have structural similarity scores >0.5, while only

~0.17% of different-fold pairs have scores >0.5. Figure 2d plots structural similarity scores between same/different-fold pairs versus their pairwise % identity. We observe an independent trend between structural similarity score and pairwise identity whereby true positives distribute to higher *structural similarity scores* (see Fig. S1 for the statistics of e-value $10^{10}$, 80% coverage threshold setting).

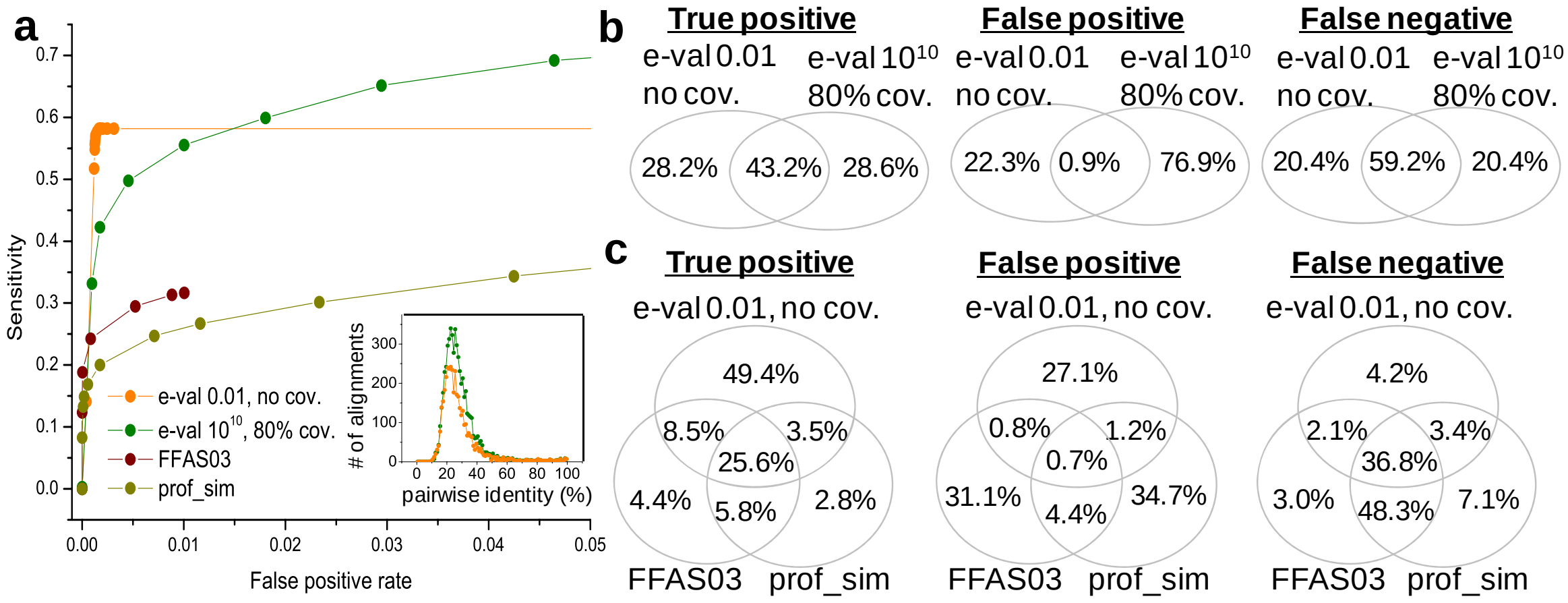

**Figure 3. Fold Recognition Performance and Comparative Statistics. (a)** Comparison of ROC curves of *structural sequence profiles* with two different settings, FFAS03 and prof_sim. Pairwise identities of the alignments between queries and the PSSMs from their true fold-specific library, which were collected with different e-value and coverage thresholds, are shown (*inset*). **(b)** Comparison of true-positive, false-positive, and false-negative pairs in top-9 result (sequences returned with the highest 9 structural similarity scores or the lowest 9 e-value/p-value for each of TZ-SABmark queries) of *structural sequence profiles* of e-value 0.01 and $10^{10}$. The numbers of true-positive, false-positive, and false-negative pairs predicted by *structural sequence profiles* of e-value 0.01 are 2645, 2161, and 4211, respectively. **(c)** Comparison of true-positive, false-positive and false-negative pairs in top-9 result of *structural sequence profiles* (e-value 0.01, no coverage), FFAS03, and prof_sim. The numbers of true-positive pairs predicted by *structural sequence profiles*, FFAS03, and prof_sim are 2666, 2030, and 1731, respectively. The numbers of false positive pairs are 613, 2776, 3075, while the numbers of true negative pairs are 4176, 4826, and 5125 (*structural sequence profiles*, FFAS03, and prof_sim respectively).

### *Performance Evaluation*

To compare our performance against other benchmarking methods, we utilized receiver operating characteristic (ROC) curve analysis(20). A ROC curve plots sensitivity versus false-positive rate, where a left-shifted curve is considered more accurate. In Figure 3a, we compare ROC curves of our method with two different settings (e-value 0.01, no coverage and e-value $10^{10}$, 80% coverage thresholds, see Fig. S2a,b for the results of different thresholds) versus two traditional fold recognition methods (FFAS03 and prof_sim(6, 8), see Fig. S2c for SAM-T2K(7)). The results demonstrate that *structural sequence profiles* of both settings outperform these benchmarking methods.

The sensitivity of *structural sequence profiles* using only statistically significant alignments from rps-BLAST (e-value 0.01, no coverage) are ~0.6 at false positive of 0.0002. *Structural sequence profiles* of e-value $10^{10}$, 80% coverage obtain similar sensitivity at a false positive rate ~0.02, but its sensitivity increases up to ~0.7 at a false positive rate 0.05 due to the additional alignments obtained. Intriguingly, the alignments obtained from both filtering strategies reside in the "twilight zone" (Fig. 3a inset). Additionally, we tested how our method performs in response to an 80% jackknife re-sampling of both queries and fold-specific libraries. The re-sampled results show a variable performance, the worst of which outperforms all other methods compared (Fig. S2d).

Figure 3b quantifies the independence between predictions of *structural sequence profiles* with two different settings (e-value 0.01, no coverage vs. e-value $10^{10}$, 80% coverage) for true-positives, false-positives, and false-negatives. Interestingly, we observe a significant number of unique true-positive pairs at both e-value settings. This suggests that comparative measurements are likely to be useful for the identification of true-positive pairs. We made the same comparison between our method (e-value 0.01, no coverage), FFAS03, and prof_sim (Fig. 3c, see Fig. S3 for comparisons using e-value $10^{10}$, 80% coverage threshold). The diagrams indicate that *structural sequence profiles* obtain more unique true-positive pairs and false-negative pairs while predicting fewer false-positive pairs. The most dramatic increase occurs between true positives whereby *structural sequence profiles* obtain 11.2 fold increase over FFAS03 and a 17.6 fold increase over prof_sim.

***Applications for Fold Classification***

Initially, we excluded PDB references sequences if they were contained in the 534 TZ-SABmark targets for building fold-specific PSSM libraries (Fig 3). To test the efficacy of *structural sequence profiles* under "real-world" conditions, we made 61 fold-specific libraries using all available PDB data. The 474 control PDB reference sequences, which are also TZ-SABmark targets, contribute on average an additional 3.6 PSSMs on average for our fold-specific libraries (Fig. S4a). As test sequences, we utilized 60 TZ-SABmark sequences which did not create PSSMs with our current settings; thus, these sequences are unrepresented in our fold-specific libraries.

In Figure 4a, we present the ROC-curve for TZ-SABmark using the complete libraries at two different thresholds (see Fig. S4b,c for different thresholds). At a false positive rate 0.001, we achieved sensitivity ~0.97 and ~0.94 for e-values of 0.01 and $10^{10}$ respectively. As expected, the pairwise % identity between TZ-SABmark test sequences and their self-generated PSSMs are low identity (~60% of the alignments are <25% identity, Fig. S5). Nevertheless, the pairwise alignments collected with both of e-value 0.01 and $10^{10}$ thresholds for encoding *structural sequence profiles* are extremely divergent (Fig. S6).

We next sought to determined if hierarchical clustering (21) of TZ-SABmark *structural sequence profiles* could rebuild appropriate SCOP fold classification (Fig. 4b). Let *accuracy* be % of TZ-SABmark queries clustered with the sequences from their respective fold groups. We calculated accuracy separately for the control and test datasets (Fig. 4c). For the control dataset, we observe 98.9% and 99.6% accuracy at e-value $10^{10}$ and 0.01 respectively. If we assign queries into the cluster with higher correlation from e-value 0.01 or $10^{10}$ (i.e. *comparative measurement*), we obtain 99.8% accuracy (e-val 0.01+e-val $10^{10}$ expected in Fig. 4c). For the test dataset, we observe 83.3%, 86.7%, and 90% accuracy at e-value $10^{10}$, 0.01, and the comparative measurement respectively. The examples which are properly clustered by comparative measurement are given in Figure S7.

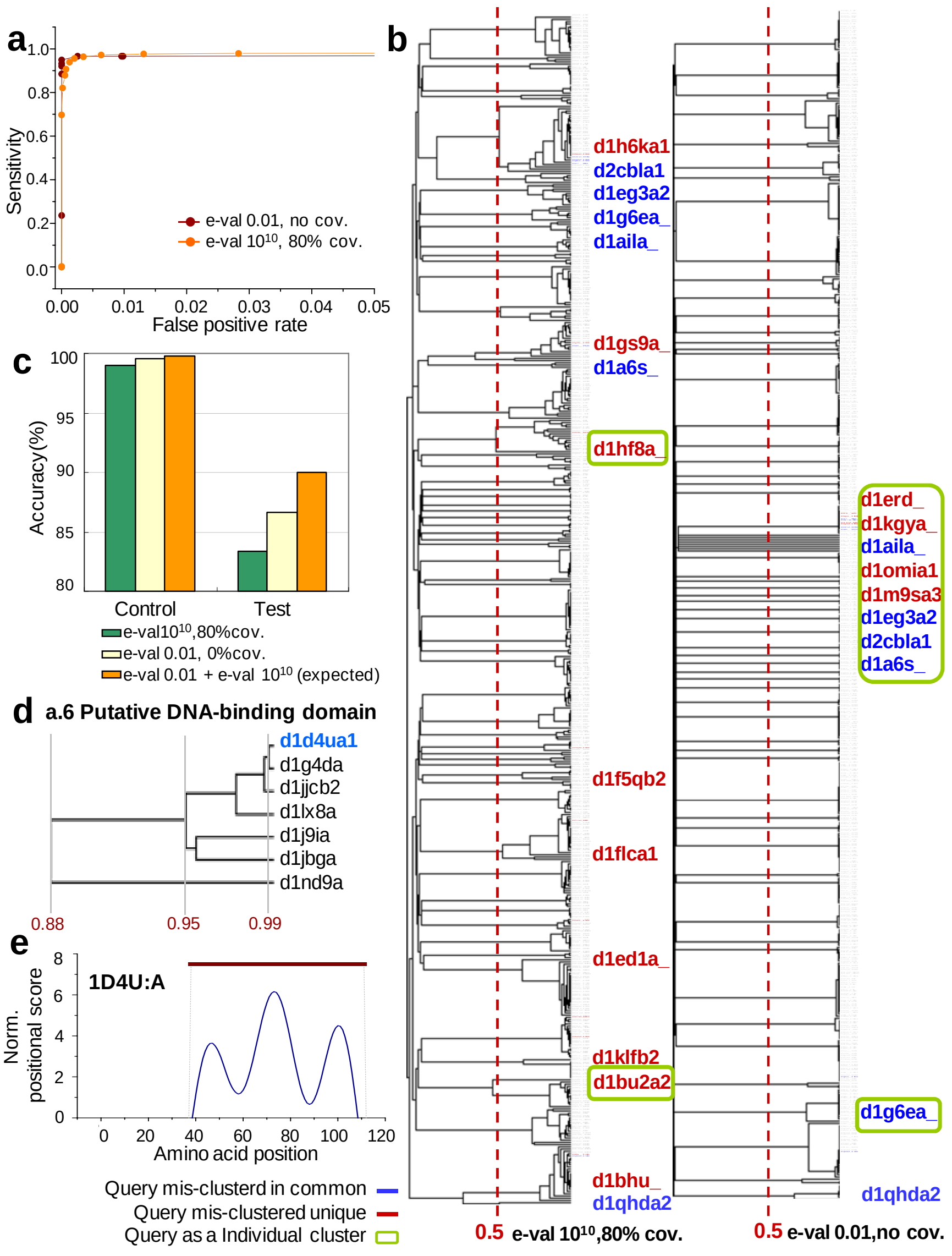

**Figure 4. Fold Recognition and Fold Clustering with the Complete Fold-specific Libraries. (a)** ROC curves for *structural sequence profiles* of e-value 0.01 and $10^{10}$ with complete fold-specific PSSM libraries. **(b)** Hierarchical clustering of structural sequence profiles of TZ-SABmark queries (*left* dendrogram: e-value $10^{10}$, 80% coverage, *right* dendrogram: e-value 0.01, no coverage threshold). The queries which could not be clustered with their related folds using Pearson's correlation 0.5 as a cutoff value (red dotted line) are in red (mis-clustered queries exclusively in either dendrogram) or blue (mis-clustered in common). The queries, which could not cluster with any other sequence as forming an individual cluster, are marked in green boxes. **(c)** Comparison of accuracy of e-value 0.01 and e-value $10^{10}$, and expected accuracy when assigning queries into the cluster with higher correlation from e-value 0.01 or $10^{10}$ (e-val 0.01 + e-val $10^{10}$ expected) for either of 474 test or 60 control TZ-SABmark queries. **(d)** By hierarchical clustering of *structural sequence profiles* (e-value $10^{10}$, 80% coverage) encoded with the complete 61 fold-specific libraries, *d1d4ua1* is correctly clustered with its true fold group without self-generated PSSMs. **(e)** Predicted a.6 SCOP fold region (blue) in the full length sequence 1D4U:A. Red line annotates an actual *Putative DNA-binding domain* (a.6) SCOP fold region in the protein. SCOP defined two domains in 1D4U:A, such as *d1d4ua1* (a.a.37-111) which is one of TZ-SABmark queries, and *d1d4ua2* (a.a.1-36). By SCOP classification, *d1d4ua1* is classified as a.6 fold while *d1d4ua2* is classified as *Glucocorticoid receptor-like* (g.39) fold. For regional prediction, ada-BLAST (10) was run with 10% seed size and 60% coverage and 10% identity thresholds using *Putative DNA-binding domain* fold-specific PSSM library.

We analyzed TZ-SABmark queries which cannot be clustered with their related fold sequences with Pearson's correlation 0.5 cutoff at e-value 0.01 and $10^{10}$ respectively (queries in red or blue in Fig 4b). Surprisingly, 9 out of the 10 queries at e-value 0.01 are not clustered with any other sequences (queries in green boxes in Fig 4b-*right*). It suggests that 99% of the TZ-SABmark queries clustered with at least a single other sequence can be accurately predicted by the fold of the sequences in the same cluster. By comparative measurement between e-value 0.01 and $10^{10}$ as previously described, we can expect 98.9% accuracy for the entire TZ-SABmark dataset because only 6 queries do not correlate with their related fold group clusters in either condition.

## DISCUSSION

In this manuscript we reveal the power of *structural sequence profiles* for fold recognition in the "twilight-zone" of sequence similarity. Our results support the hypothesis that *structural sequence profiles* provide a viable solution to the inverse protein folding problem. This is supported by several key findings from our measurements: (i) "twilight-zone" pairwise alignments are informative (Fig. 2), (ii) they outperform multiple benchmarking methods in TZ-SABmark by providing more unique true-positive pairs (Fig. 3), and (iii) they are capable of reconstituting structural fold classifications (Fig. 4). A number of broad implications can be derived from this study.

We previously theorized that low-identity alignments are a rich source of information, which can be used to unmask the fundamental properties of proteins, including protein structure, function, and evolution using simple arithmetic(9, 10, 15). As described above we build diverse PSSM libraries which are scalable using PSI-BLAST and searchable with rps-BLAST. We take advantage of the information content provided by PSSMs to increase the signal-to-noise ratio inherent to low-identity alignments. In addition, we demonstrate that a coverage threshold is an effective filter of noisy alignments (Fig S2a, S4b). When fold-specific scores are encoded into a vector (i.e., *structural sequence profiles*), multiple data mining algorithms can be used reliably to measure fold attributes.

We also evaluated the performance of *structural sequence profiles* correlated using Pearson's correlation coefficients to relate divergent structural folds. When compared to popular profile-based algorithms such as FFAS03, SAM-T2K, and prof_sim, *structural sequence profiles* obtain a significant portion of unique true-positive pairs and reduced false-positives. Taken together, this underlies our increased performance. Interestingly, all methods including *structural sequence profiles* encoded with the alignments obtained at different thresholds recover a substantial number of unique pairs; however, relating these unique pairs (outside of our own) is difficult as each method provides data which is not necessarily interoperable.

Considering the current genomic explosion of sequences, fold-recognition methods are needed as they are a true watershed in Biology. Based on the results presented here, conversion and PSI-BLAST expansion of the PDB into fold-, superfamily-, and family-specific PSSM libraries would, in theory, synergize and improve structural modeling in general. For example, given the correct fold and nearest structurally resolved neighbor for d1d4ua1 (Fig. 4d), threading algorithms could be employed to generate a relevant model. In addition, for physical-based and hybrid modeling approaches, these same data would significantly reduce the folding space, which would likely improve the efficacy, speed, and refinement capable by these algorithms. In closing, we propose that encoding all known folds into fold-specific libraries should enable a comprehensive assignment of structural folds for a substantial portion of all known and yet to be discovered sequences.

## MATERIALS AND METHODS

**Benchmarking methods.** SAM-T2K, prof_sim, and FFAS03 are used as benchmark methods (6-8). For SAM-T2K, *blastall* in NCBI BLAST 2.2.15 is used for *target2k* script in SAM3.5 package searching a sequence database to collect sequences for HMM generations for 534 test sequences. When a query sequence is scored given a HMM model by *hmmscore*, Smith-Waterman algorithm was used by default. For prof_sim, sequence profiles were generated by PSI-BLAST and profile-profile alignment was done with local alignment setting. For all of three benchmark methods, NCBI NR database with 6,419,591 protein sequences was used as a sequence database. FFAS03 was run by a member of Gozik lab(6) to a false-positive rate ~0.01. In the result of each method, all-against-all comparison of TZ-SABmark test sequences was done, and for each sequence, all other sequences are sorted by e-value or p-value for ROC curve analysis.

**Fold-specific library construction.** To build fold-specific PSSM libraries, reference PDB sequences were collected. Using each of the PDB sequences as a query, a PSI-BLAST search was performed against NCBI NR database. The settings for PSI-BLAST search were 3 maximum number of iterations, 30 maximum number of database sequences returned at each iteration (-b option), and other options remained as default. Among the returned database sequences, the sequences with more than 90% sequence identity to the query PDB query sequence were removed. Further, a Needleman-Wunsch global alignment was performed between pairs of the sequences(22), and, for the pairs with >=40% identity, one was removed. PSSMs are generated for the collected sequences, after filtering, to build each fold-specific PSSM library by PSI-BLAST (6 iterations, e-value $10^{-6}$). PSSM libraries and scripts are available upon request.

**Fold-specific score.** Given the alignments returned from the rps-BLAST search of each query against a fold-specific PSSM library, each amino acid of a query is scored +2 for identically aligned residues and +1 for positively aligned residues (not identical, but conserved). These results are summed for each amino acid of the query and normalized by subtracting the average positional score of all amino acids. The score is calculated as: $\frac{1}{n}\sum_{i=1}^{n} r_i$ if $r_i > 0$ where n is the length of a protein sequence and $r_i$ is residue score of i^th^ amino acid of the protein.

**Pearson's correlation coefficient.** The structural similarity score of two proteins is the Pearson's correlation coefficient between their structural sequence profiles X and Y, *PC(X, Y),* is calculated as: $PC(X,Y) = \frac{1}{n}\sum_{i=1}^{n}\left(\frac{X_u - \mu_X}{\delta_X}\right)\left(\frac{Y_i - \mu_Y}{\delta_Y}\right)$ where $n$ is the number of measuring folds, and $\mu_X$ and $\delta_X$ are the average and standard deviation of X.

**ROC curve analysis.** To calculate sensitivity at different false-positive rates for each query, the sequences with higher structural similarity score (in case of our method) or smaller e-value/p-value (in case of benchmarking methods) than different cutoff thresholds are returned as homologues by each method. Sensitivity and false-positive rates are calculated as: Sensitivity $=\frac{TP}{TP+TN}$, false positive rate $=1-\frac{TN}{TN+FP}$, when TP = the number of true positives, TN = the number of true negatives, and FP = the number of false positives.

**Jacknife resampling.** To estimate the variance of our performance, 80% jackknife resampling was performed. 10 sample datasets were generated with TZ-SABmark queries of randomly selected 80% test fold groups. For the queries of each sample dataset, *structural sequence profiles* were generated with the alignments between the queries and randomly selected 80% fold-specific libraries.

**Fold hierarchy.** To build a fold classification hierarchy, hierarchical clustering with complete linkage was performed given structural sequence profiles of TZ-SABmark queries using Pearson's correlation coefficient as similarity metric. For clustering and dendrogram visualization, Cluster 3.0 and Java Treeview 1.14 were utilized.

**ACKNOWLEDGEMENTS**

This work was supported by the Searle Young Investigators Award and start-up monies from Pennsylvania State University (R.L.P.), Funds from the Huck Life Science Institute's Center for Computational Proteomics (R.L.P and D.V.R) and a grant from the Pennsylvania Department of Health using Tobacco Settlement Funds to D.V.R. The Pennsylvania Department of Health specifically disclaims responsibility for any analyses, interpretations, or conclusions. This work was also supported by The National Science Foundation 428-15 691M (RLP, DVR), and The National Institutes of Health R01 GM087410-01 (RLP, DVR). We would like to thank the CAC Center and Jason Holmes for their exceptional support, and to the Gozik lab for running our dataset using FFAS30. Additionally we would like to thank Dr. Jayanth Banavar for inspiring this project, and Drs. Robert E. Rothe, Jim White, Max Moon, Peter J. Hudson, Dee Lite, Gene and Dean Ween, Barbara VanRossum, Anderson R. Sharer, and Jesus Silliman for creative dialogue.

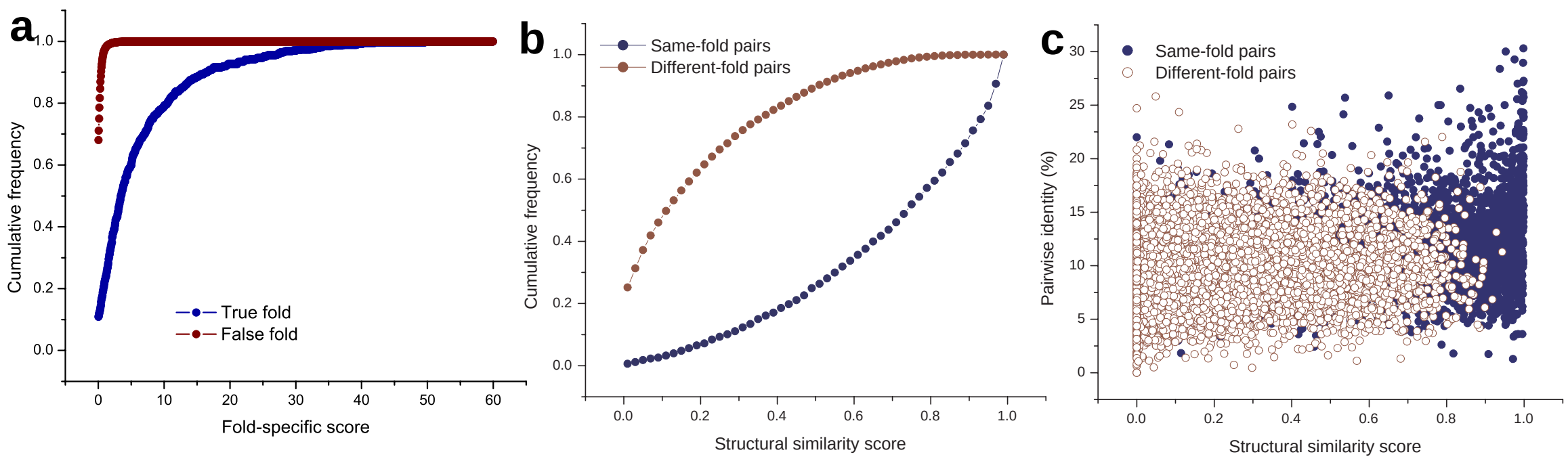

**Supplemental Figure 1. Characterization of Structural Similarity Scores given e-value 1010 and 80% coverage threshold.** (a) The distributions of query sequence scores for each fold-specific library. 71.1% of the query sequences have fold-specific scores 0.1 for different-folds, while only 11.1% of them have scores 0.1 for same-folds. (b) Cumulative frequencies of the structural similarity scores between pairs of same-fold (blue) and different-fold (red) query sequences. 75.0% of same-fold pairs have structural similarity scores >0.5, while only 16.1% of different-fold pairs have scores >0.5. For this measurement, 3,428 same-fold pairs and 65,536 different-fold pairs were measured from 534 sequences. (c) Structural similarity scores between pairs of same-fold and different-fold query sequences were plotted versus their pairwise sequence identities. This data shows an independent trend between the structural similarity score and pairwise identity in the “twilight-zone” of sequence similarity. The data points of randomly selected 10,000 different-fold pairs were plotted.

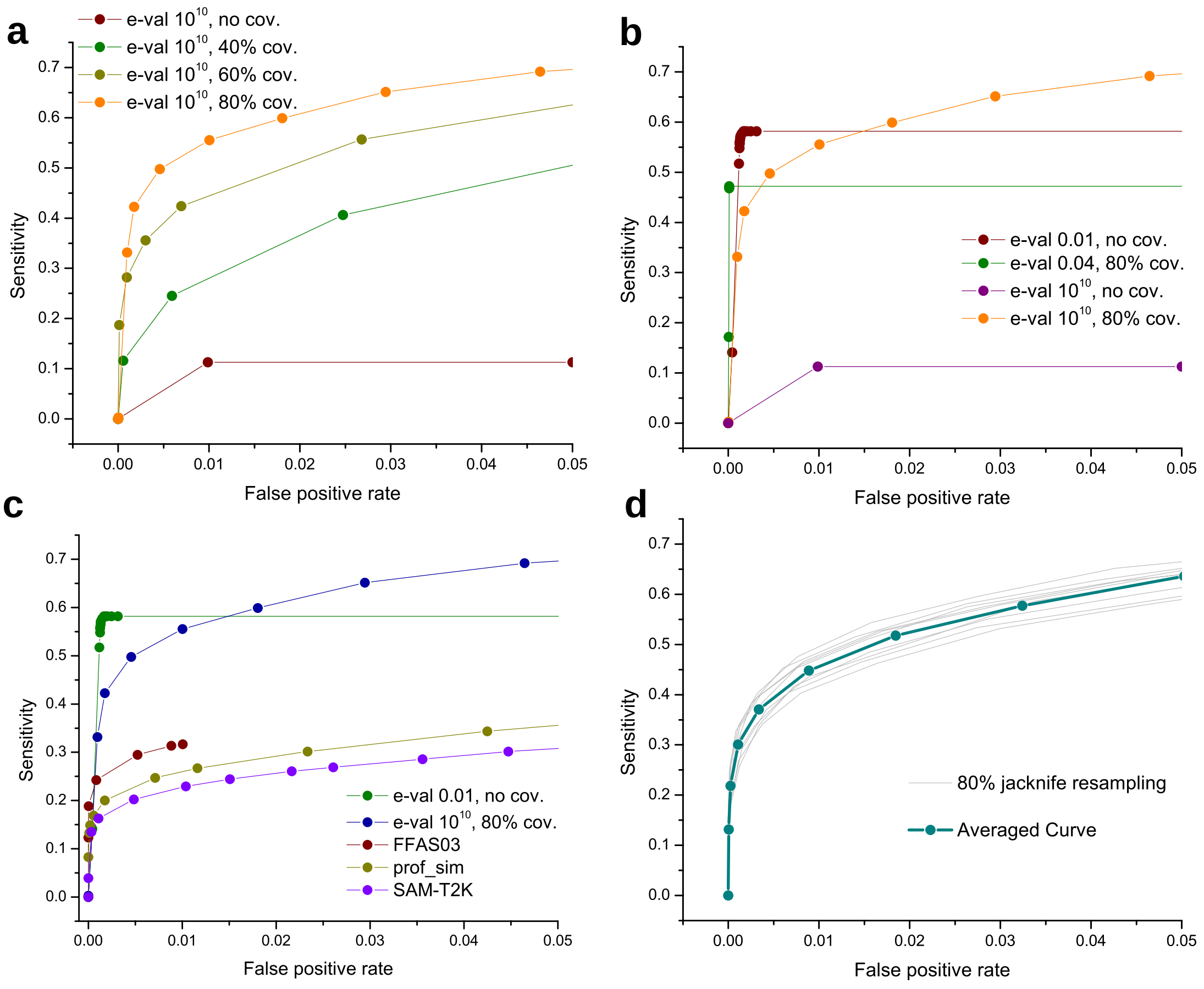

**Supplemental Figure 2. Fold Recognition Performance of *Structural Sequence Profiles* with Different Settings Given 61 Fold-specific Libraries of 6,470 PSSMs without self-generated PSSMs. (a)** Comparison of ROC curves of *structural sequence profiles* at different coverage thresholds when e-value threshold is fixed at $10^{10}$ **(b)** Comparison of ROC curves of *structural sequence profiles* of e-value 0.01 & no coverage & e-value 0.01, 80% coverage, e-value $10^{10}$ & no coverage, and e-value $10^{10}$ & 80% coverage. **(c)** Comparison of ROC curves of *structural sequence profiles* with two different settings (of e-value 0.01, no coverage and e-value $10^{10}$, 80% coverage), FFAS03, prof_sim, and SAM-T2K. **(d)** ROC curves of *structural sequence profiles* of e-value 0.01, 80% coverage given 10 random samples of queries and fold-specific libraries re-sampled by 80% jackknifing. Each grey curve is ROC curve of each of 10 random samples. Blue curve is the average curve of the 10 ROC curves. In these cases, there is no guarantee that we will measure the query sequences with their true-fold libraries. Thus, if two proteins are structurally similar, we reason that they would have similar pattern of scores for different-fold libraries, and such similarity would lead our method to see their structural similarity albeit indirectly and with a reduced efficiency.

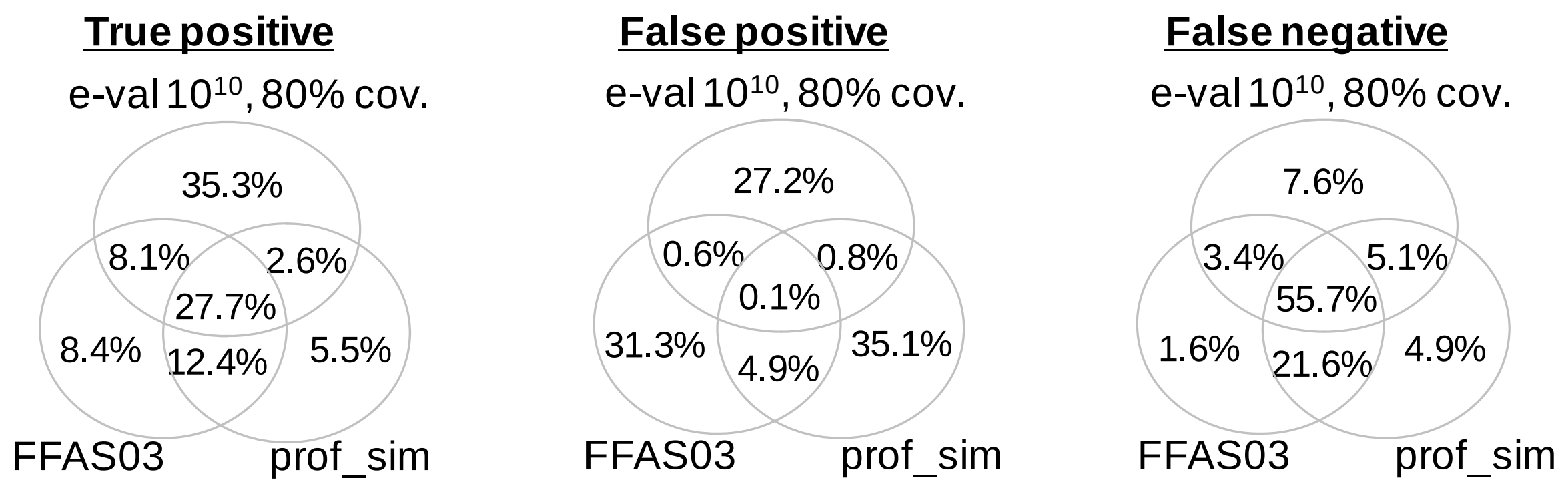

**Supplemental Figure 3. Comparison of true-positive, false-positive and false-negative pairs in top-9 result of structural sequence profiles of e-value $10^{10}$, 80% coverage threshold, FFAS03, and prof_sim.** The numbers of true-positive pairs predicted by structural sequence profiles, FFAS03, and prof_sim are 2,645, 2,030, and 1,731, respectively. The numbers of false positive pairs are 2,161, 2,776, 3,075, while the numbers of true negative pairs are 4211, 4826, and 5125 (*structural sequence profiles*, FFAS03, and prof_sim respectively).

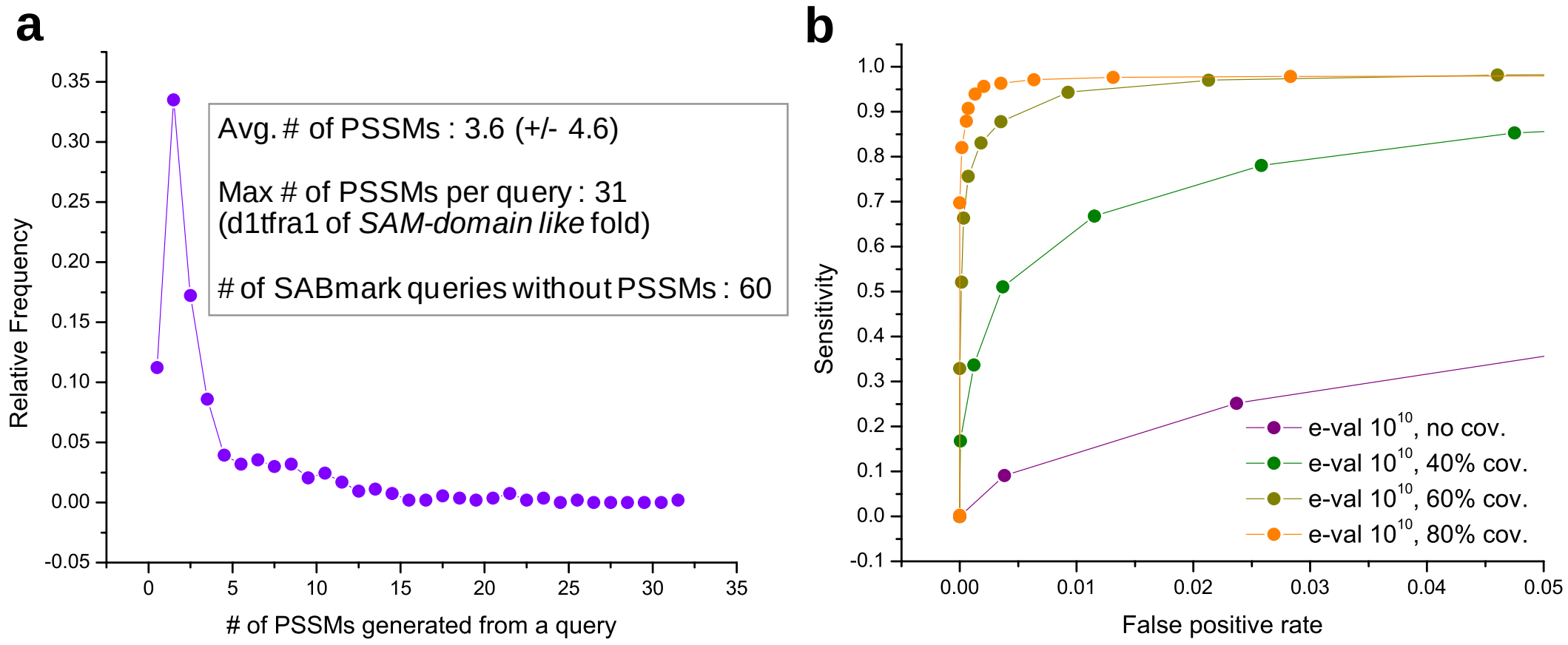

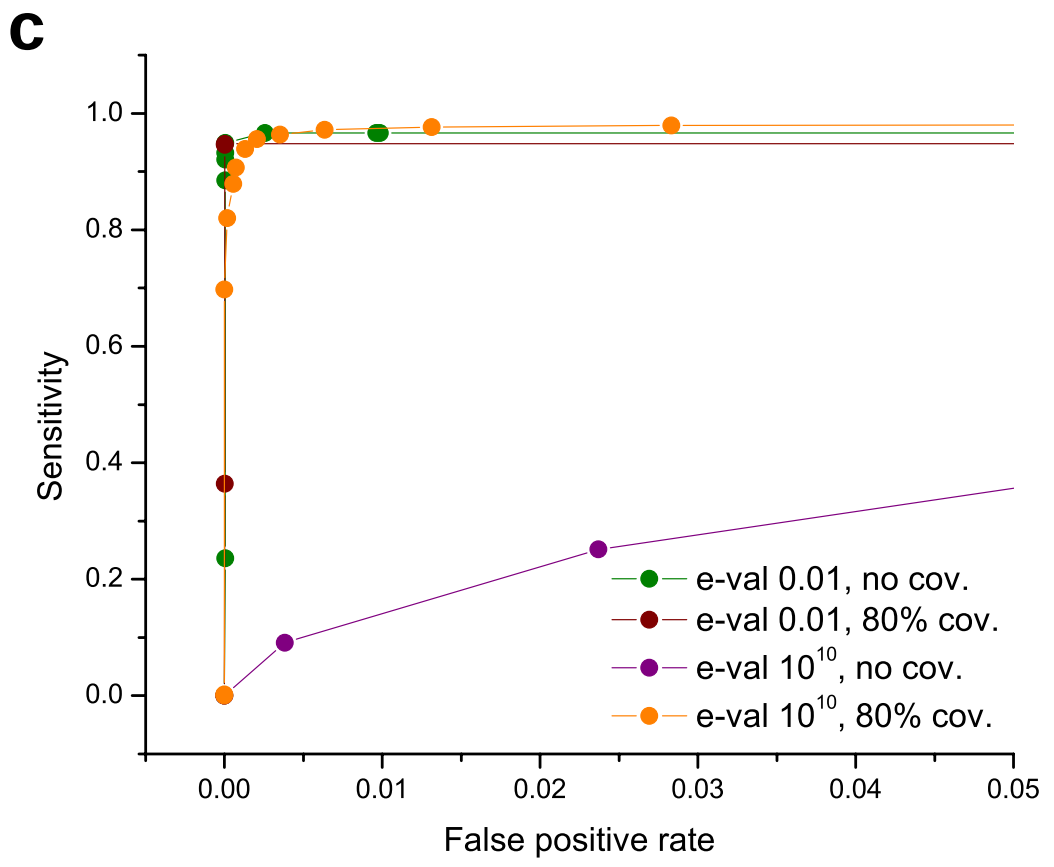

**Supplemental Figure 4. Fold Recognition Performance of *Structural Sequence Profiles* with Different Settings Given the Complete 61 Fold-specific Libraries of 8,359 PSSMs Including Self-generated PSSMs.** **(a)** Relative frequency of # of PSSMs generated from each TZ-SABmark queries. The average number of PSSMs generated from the queries is 3.6, with the most being 31 PSSMs which were generated from *d1tfra1*, a *SAM-domain like* fold. If a TZ-SABmark query was not in the version of PDB which we used to construct fold-specific libraries, PSSMs were not generated from the query. There are 60 such TZ-SABmark queries. **(b)** Comparison of ROC curves of *structural sequence profiles* at different coverage thresholds when e-value threshold is fixed at $10^{10}$. **(c)** Comparison of ROC curves of *structural sequence profiles* of e-value 0.01 & no coverage, e-value 0.01 & 80% coverage, e-value $10^{10}$ & no coverage, and e-value $10^{10}$ & 80% coverage.

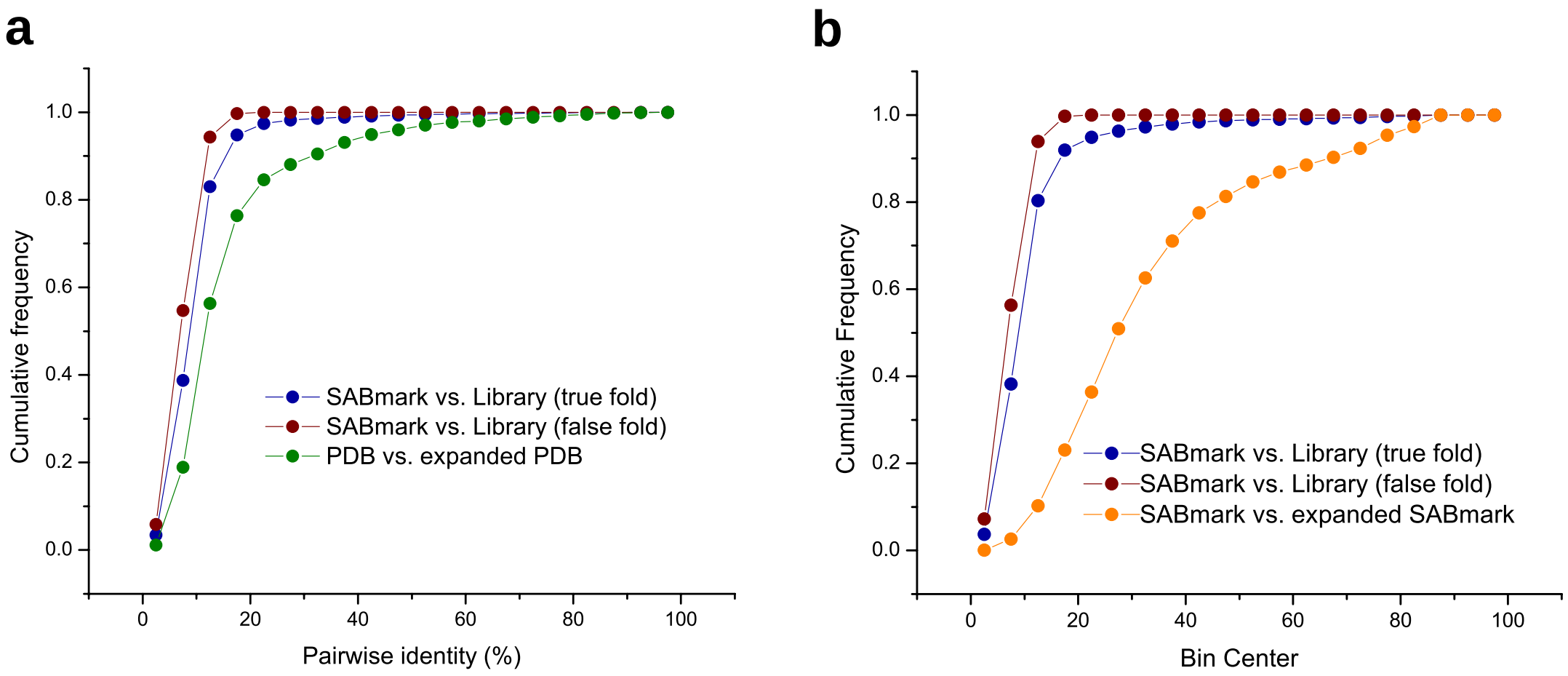

**Supplemental Figure 5. Characterization of Sequence Similarity Between TZ-SABmark Queries and Library Sequences.** **(a)** The distribution of % pairwise identity between pairs of TZ-SABmark query sequences and the library sequences of the same-fold (blue) and different-folds (red). Pairwise identity between the original PDB sequences and PSI-BLAST expanded sequences (green). The pairwise identity was calculated from Needleman-Wunsch global alignments with BLOSUM62, Gap opening penalty 10, and Gap extension penalty 0.5. **(b)** The distribution of %pairwise identity between pairs of TZ-SABmark query sequences and the library sequences generated directly from the queries (orange), along with distribution of %pairwise identity between pairs of TZ-SABmark query sequences and the library sequences of the same-fold (blue) and different-folds (red) shown in (a).

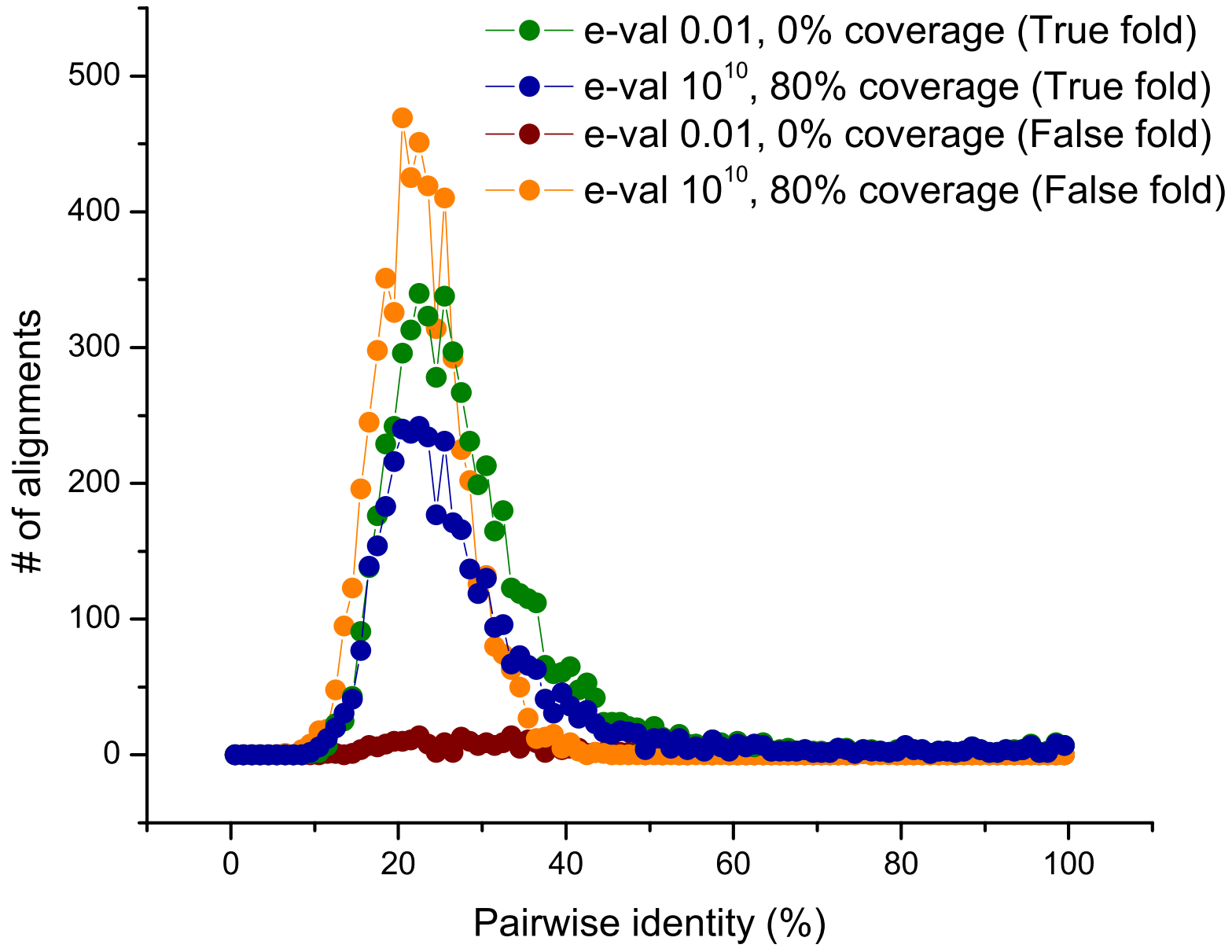

**Supplemental Figure 6. Characterization of alignments encoded in *structural sequence profiles* of e-value 0.01 and $10^{10}$ thresholds.** Pairwise identity of the alignments collected by rps-BLAST with e-value 0.01, no coverage and e-value $10^{10}$, 80% coverage for *structural sequence profile* encoding were analyzed. The analysis is done separately for the alignments between queries and the PSSMs of true fold-specific library and those between the queries and the PSSMs of false fold-specific library.

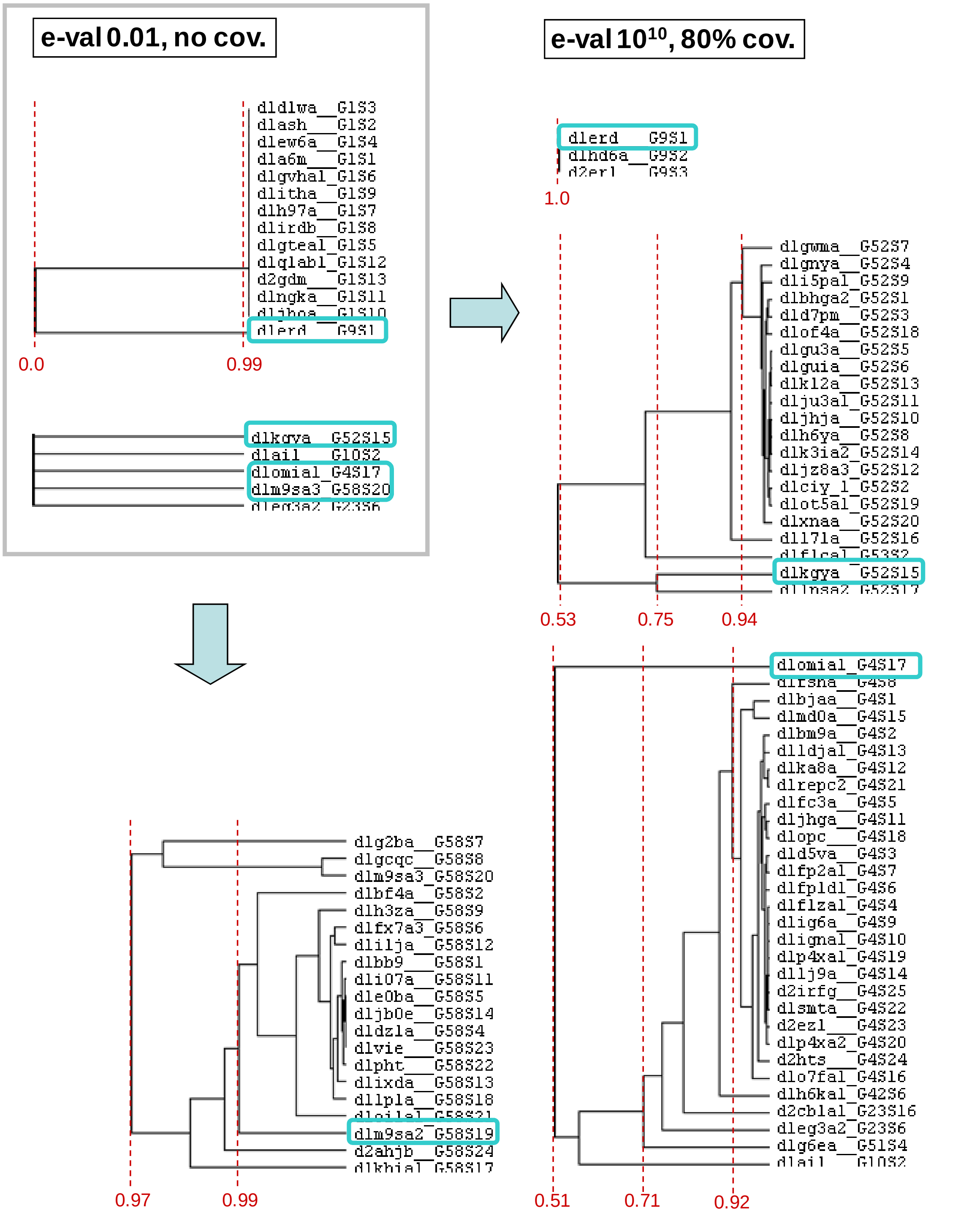

**Supplemental Figure 7. Comparison of *structural sequence profile* dendrograms of e-value 0.01 and $10^{10}$ thresholds.** A portion of the e-value 0.01, no coverage dendrogram containing the queries mis-clustered (*top left box)* and a portion of the e-value $10^{10}$, 80% coverage dendrogram containing the same queries. The queries in blue boxes (queries in red in the dendrogram of e-value 0.01, no coverage in Fig. 4b) were improperly clustered in the result of e-value 0.01, no coverage, but properly clustered in the result of e-value $10^{10}$, 80% coverage.